\documentclass[12pt,osajnl2,preprint]{revtex4}  
\usepackage{graphics}
\usepackage{graphicx}

\newcommand*{\cm}{cm$^{-1}$}

\begin{document}

\title{Near-infrared studies of glucose and sucrose in aqueous solutions: water displacement effect and red shift in water absorption from water-solute interaction}

\author{Youngeui Jung$^{1}$ and Jungseek Hwang$^{1,2,*}$}
\address{$^{1}$Department of Physics, Pusan National University, Busan 609-735, Republic of Korea}
\address{$^{2}$Department of Physics, Sungkyunkwan University, Gyeonggi-do, Suwon 440-746, Republic of Korea}

\address{$^*$Corresponding author: jungseek@skku.edu}

\begin{abstract}
We use near infrared spectroscopy to obtain concentration dependent glucose absorption spectra in their aqueous solutions in the near-infrared range (3800 - 7500 cm$^{-1}$). We introduce a new method to obtain reliable glucose absorption bands from aqueous glucose solutions without measuring the water displacement coefficients of glucose separately. Additionally, we are able to extract the water displacement coefficients of glucose, and this may give a new general method using spectroscopy techniques applicable to other water soluble materials. We also observe red shifts in the absorption bands of water in the hydration shell around solute molecules, which comes from contribution of the interacting water molecules around the glucose molecules in solutions. The intensity of the red shift get larger as the concentration increases, which indicates that as the concentration increases more water molecules are involved in the interaction. However, the red shift in frequency does not seem to depend significantly on the concentration up to our highest concentration. We also performed the same measurements and analysis with sucrose instead of glucose as solute and compare.
\end{abstract}


\maketitle 

{\noindent}Index Headings: Near infrared spectroscopy, Water displacement coefficient, Glucose solution, Sucrose solution\\

{\noindent}\textbf{INTRODUCTION}

The possibility of providing a direct, non-invasive approach to measuring glucose concentrations in blood has inspired studies of the applications of infrared spectroscopy for analyte detection in various solutions. It has been shown that glucose has many distinct infrared (IR) absorption features in the far-infrared (FIR),\cite{upadhya04, upadhya03, walther03} mid-infrared (MIR),\cite{lowying02, petibois99, jenson04, shen03,klonoff98} and near-infrared (NIR)\cite{heise96,riley97,hazen98,riley00,malchoff02,amerov04,he11} regions. However, water, the main component of blood, also displays strong IR absorption features in these regions, with increased absorption as you move further towards FIR. Water absorption modes in a wide spectral range can be found in a literature.\cite{tennyson01}

In studies dealing with MIR analyte detection in blood,\cite{lowying02, petibois99} the experiments were performed on dry samples or using second derivative spectra to obtain concentration dependence.\cite{shen03} This causes a large inaccuracies due to an increase in noise because of derivation. Water IR absorption in the MIR region as well as the FIR is overwhelming,\cite{matei03} rendering non-invasive glucose detection at physiological concentrations extremely difficult due to high water content in blood. In a recent work,\cite{amerov04} glucose absorption bands were extracted by using an independently measured water displacement coefficient of glucose. The water displacement coefficient is a measure of the second order effect of the presence of glucose on the spectrum of water. Since, at physiologically relevant concentration of glucose in blood, the water bands are orders of magnitude stronger than the glucose bands, incorrect treatment can be a major source of error.

In this study we introduce a new method to obtain reliable glucose absorption bands without measuring the water displacement coefficient separately. Because there is a strong water absorption peak at 5200 cm$^{-1}$ and almost no or negligible glucose absorption at this frequency, we take advantage of this strong and isolated water absorption peak to remove water absorption from the measured transmission of aqueous glucose solutions. By adjusting effective thicknesses of water in a liquid cell to match the amplitude of the water peak for six different glucose solutions in the same liquid cell we are able to remove water absorption bands accurately and to extract an accurate concentration dependent glucose absorption coefficient in solution. Additionally, we are able to obtain a water displacement coefficients of glucose by using the concentration dependent effective thickness of water in the cell. We also observe red shifts in the absorption bands of water in the solution. Similar there are red shifts from interacting water in macroscopic air-water and oil-water interfaces and in the hydration cell around nonpolar hydrocarbon solute groups.\cite{moore06,perera09} Water structure enhancement within hydration shells was reported.\cite{hvidt83,raschke05} This indicates that water molecules around solute molecules are not free; they are interacting with solute molecules. Our analysis shows that the number of interacting water molecules seems to increase as the concentration of solute increases. We also apply the same method to another water soluble material, sucrose, which has a higher molecular weight than glucose. We compare the results of sucrose with those of glucose.\\

{\noindent}\textbf{EXPERIMENTS}

We prepared six different aqueous glucose and six sucrose solutions: the concentrations were 1.00, 2.00, 4.00, 6.00, 8.00 and 10.00 g/dL. All solutions were prepared using anhydrous D-(+)glucose (C$_6$H$_{12}$O$_6$) purchased from Sigma-Aldrich (USA), sucrose (C$_{12}$H$_{22}$O$_{11}$) purchased from Junsei Chemical (Japan), and grade-3 deionized water. Aqueous samples were placed in a 250 $\pm 10\ \mu$m path-length liquid cell composed of glass. The cell was made using epoxy glue to attach 155 $\mu$m thick pieces of microscope cover glass (Sargent-Welch, USA) to a 1-mm thick microscope slide (VWR Scientific, USA) forming a rectangular chamber ($\sim$0.250$\times$11.2$\times$17.5 mm$^3$). By covering this with another microscope slide we created a 250 $\pm 10\ \mu$m thick liquid cell appropriate for aqueous sample measurements. Reproducibility and stability of the measurement system were tested before proceeding with the study. We also prepared a pure amorphous glucose pellet and a sucrose pellet melting the D-glucose and sucrose powders, respectively. We measured those pellets to obtain the absorption coefficients of pure glucose and sucrose.

A commercial Fourier transform infrared (FTIR) spectrometer, Bruker Vertex 80v was used for collecting near infrared spectra. The optical setup consists of a 75 W tungsten lamp as a light source, a CaF$_2$ beam splitter, and a room temperature deuterated triglycine sulfate (DTGS) detector. We measured transmittance spectra, $T(\omega)$, of samples on a sample holder with a 5.0 mm diameter circular aperture and a resolution of 5 \cm\ over a range of 3800 - 7500 \cm. An empty glass cell was used as the reference for all transmittance measurements except for the amorphous glucose and sucrose pellets. To get transmittances of the two pellets an empty hole was used as the reference. We note that for all transmittance measurements of liquid samples we used the same liquid cell. All transmittance spectra were taken at room temperature (23 $^{o}$C). Absorption coefficient spectra, $\alpha(\omega)$, were calculated from measured transmittances.\\

{\noindent}\textbf{MEASURED DATA AND ANALYSIS}

\begin{figure}[h]
  \centering\includegraphics[width=4.5 in]{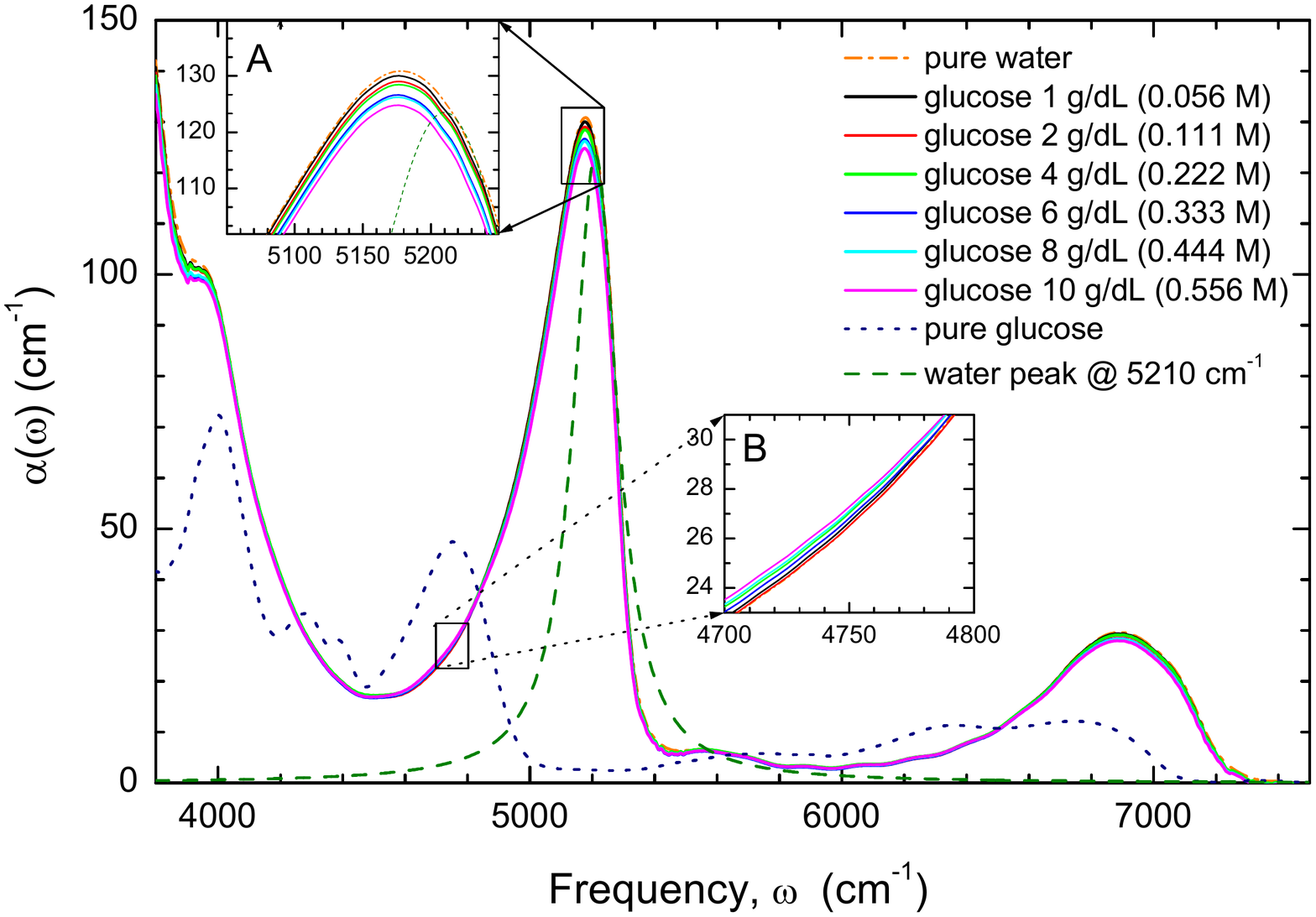}
  \caption{(Color online) Absorption coefficients of a pure amorphous glucose (dotted dark blue line) a pure water (dash-dotted orange line), and six glucose aqueous solutions (from top to bottom; from low to high concentrations). We also show a water peak at 5210 \cm\ (green dashed line). In the inset A and B we show expanded views near 5150 \cm\ and 4750 \cm\ respectively.}
  \label{fig1a}
\end{figure}

We measured transmittance spectra of the six different glucose and six sucrose solutions and pure water in the cell as well as pure amorphous glucose and sucrose pellets. The absorption coefficient can be extracted from a measured transmittance spectrum by using the well-known Beer-Lambert's formula:
\begin{equation}\label{eq1}
\alpha(\omega, C)=-\frac{\ln T(\omega, C)}{d}
\end{equation}
where $\alpha(\omega)$ is the absorption coefficient, $T(\omega)$ is the measured transmittance, $C$ is the concentration of the solution and $d$ is the thickness of the sample. Figure \ref{fig1a} shows raw absorption coefficients for the six glucose solutions, pure water, and a pure amorphous glucose pellet. For the solution samples (water and solutions) we used the same thickness $d_0\cong$ 252 $\mu$m since we used the same liquid cell. We also show a water absorption peak at 5200 \cm, which is a Lorentzian function. Since glucose absorption in the solution is very weak compared with the water absorption at these concentrations, we can not see large differences among in solution spectra. In these solution spectra, we observe three strong water absorption peaks in a spectral range between 3800 and 7500 \cm, which are near 4000, 5200, and 6900 \cm. There are two physiologically relevant windows in the water absorption through this measured spectral range; one between 4000 \cm\ and 5200 \cm\ is the combination region where four distinct glucose peaks are visible and the other between 5200 \cm\ and 6900 \cm\ is the first overtone region where two broad glucose peaks which are not as distinct or strongly absorbing as those in the combination region.

\begin{figure}[h]
  \centering\includegraphics[width=3.5 in]{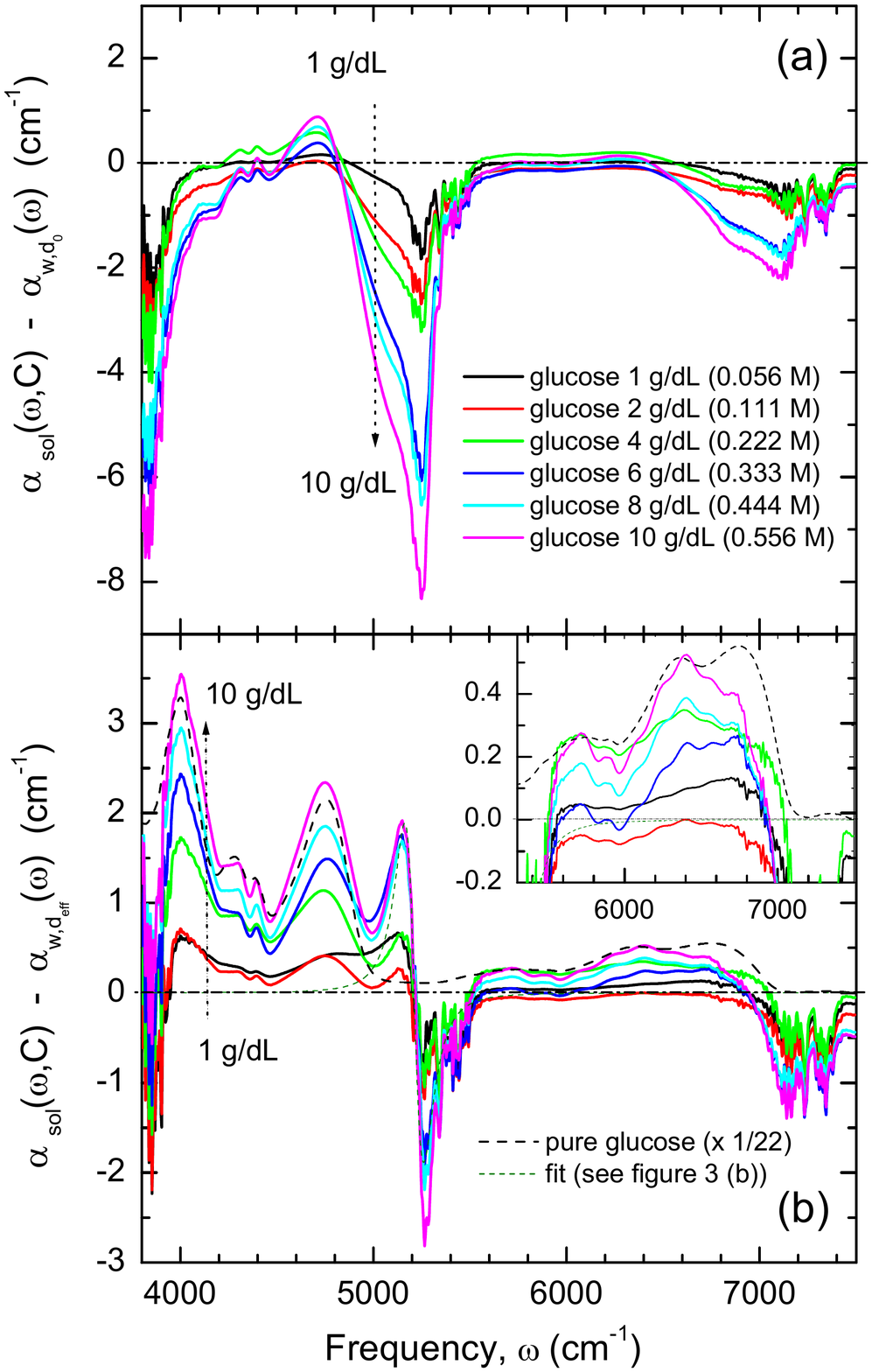}
  \caption{(Color online) (a) Glucose absorption bands in solutions obtained at six different concentrations by subtracting water spectrum from those of the glucose solutions. Water displacement coefficient of glucose has not been considered on these spectra (see in the text). (b) Glucose absorption coefficients in six different solutions. Water displacement effects have been taken into account for the subtracting water absorption procedure (see in the text). The green dotted line is a fitted line to the feature near 5200 \cm with both peak and dip from a model calculation (see figure \ref{fig2} (b) and corresponding text). The black dashed line in the lower frame is the pure glucose absorption spectra with its intensity reduced by a factor of 22. In the inset we display an expanded view to show spectral features better in the first overtone region.}
  \label{fig1b}
\end{figure}

To obtain the absolute magnitude of glucose absorption from a glucose solution, we initially subtracted the measured water absorption coefficient from that of each solution. As a first approximation, we assumed that the thicknesses of water are the same. Then we can formulate the subtraction procedure as follows:
\begin{equation}\label{eq1a}
\alpha_{sol}(\omega, C)-\alpha_{w,d_0}(\omega)=-\frac{\ln [T_{sol}(\omega, C)]}{d_0}-\Big{[}-\frac{\ln[ T_{w}(\omega)]}{d_0}\Big{]}
\end{equation}
where $d_0$ is the thickness of our liquid cell. $\alpha_{sol}(\omega)$ and $\alpha_{w,d_0}(\omega)$ are respectively the absorption coefficients of a solution and pure water calculated using the cell thickness $d_0$ = 252 $\mu$m. $T_{sol}(\omega)$ and $T_{w}(\omega)$ are the measured transmittance spectra of the solution and pure water, respectively. Figure \ref{fig1b} (a) shows spectra resulting from this analysis procedure. There is only one clearly visible glucose peak at 4700 \cm, offset from the actual peak position seen in figure \ref{fig1a} of 4740 \cm. Also, at 5200 \cm, there is a sharp downward peak, with the larger peak for the higher concentrated solution. This concentration dependent downward peak appears in the difference spectra because we did not consider the water displacement effect due to the glucose presence in the solution. When glucose is dissolved in water the volume of the solution changes because each glucose molecule takes up a finite space. We have to take into account this (water displacement) effect to subtract an appropriate water spectrum from the solution spectra.

\begin{figure}[h]
  \centering\includegraphics[width=3.5 in]{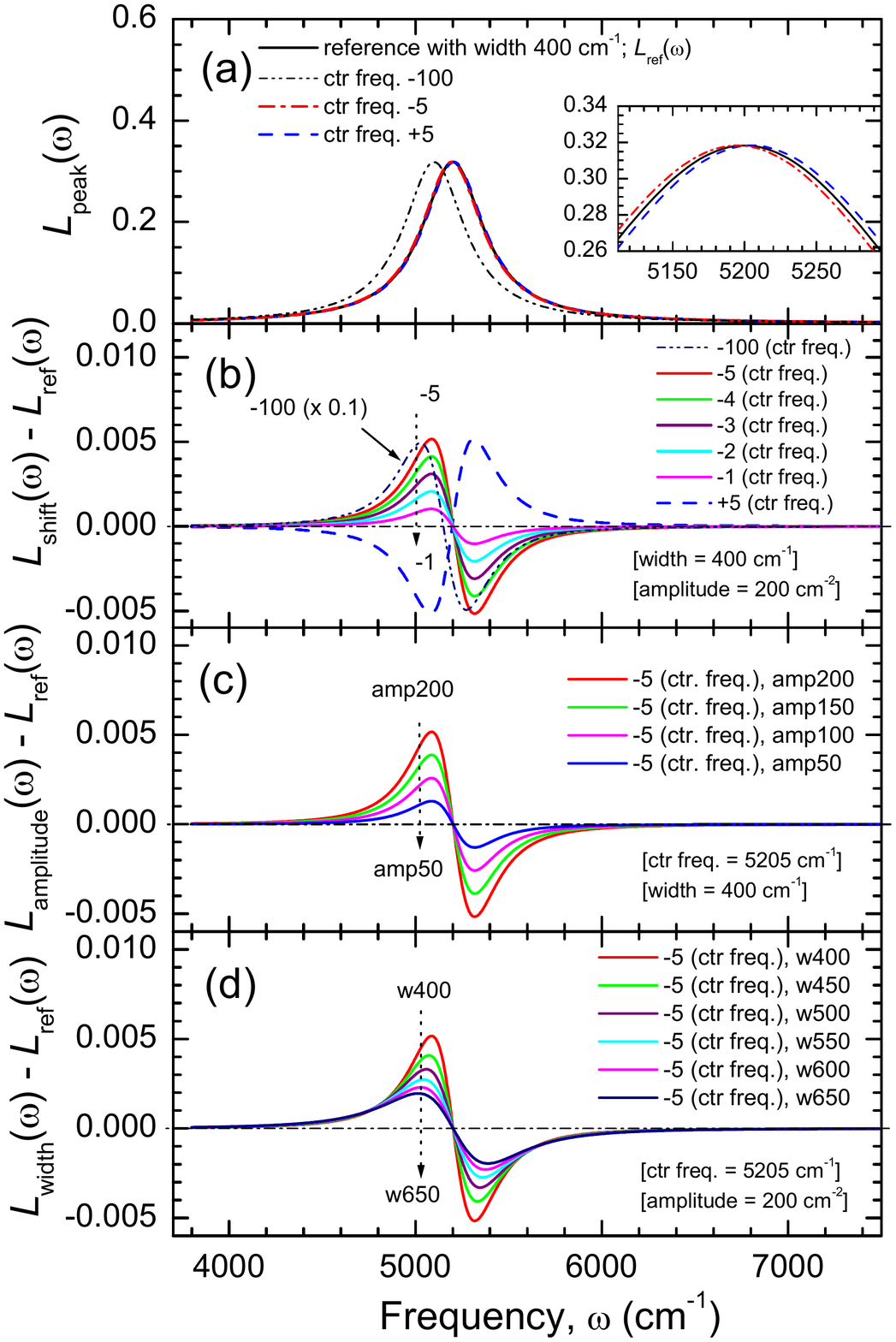}
  \caption{(Color online) (a) We show a reference peak which has its center at 5200 \cm\ and width 400 \cm. We also show two shifted peaks in horizontal direction by negative 5 \cm\ (dashed dotted red curve) and positive 5 \cm\ (dashed blue curve), respectively. In the inset we expand the graphs near the peak region to show the shifts more clearly. (b) We also show resulting differences subtracted the reference peak at 5200 \cm\ from the shifted peaks by seven different amounts  (see in the text). (c) We show resulting differences subtracted the reference peak at 5200 \cm\ from the peaks at 5205 \cm\ with four different amplitudes (see in the text). (d) (c) We show resulting differences subtracted the reference peak at 5200 \cm\ from the peaks at 5205 \cm\ with six different widths (see in the text).}
  \label{fig2}
\end{figure}

\begin{figure}[h]
  \centering\includegraphics[width=4.5 in]{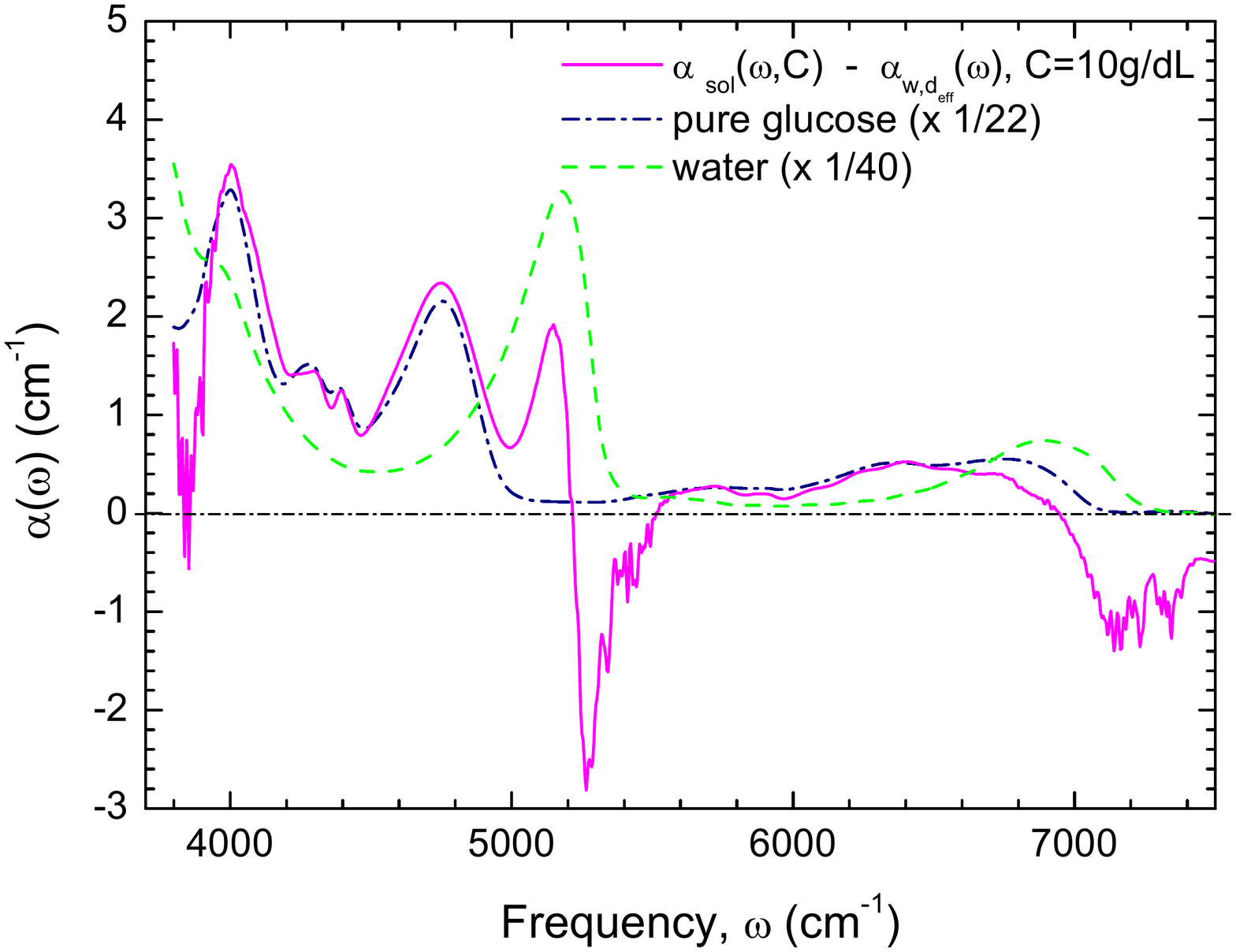}
  \caption{(Color online) Comparison the pure glucose absorption and the extracted glucose absorption ($\alpha_{sol}(\omega, C)-\alpha_{w,d_{eff}}(\omega)$) from 10g/dL solution. We note that we reduce the intensity of the pure glucose absorption by a factor of 22. We also show the water spectrum as well for comparison purpose and reduce its intensity by a factor of 40.}
  \label{fig3a}
\end{figure}

\begin{figure}[h]
  \centering\includegraphics[width=3.5 in]{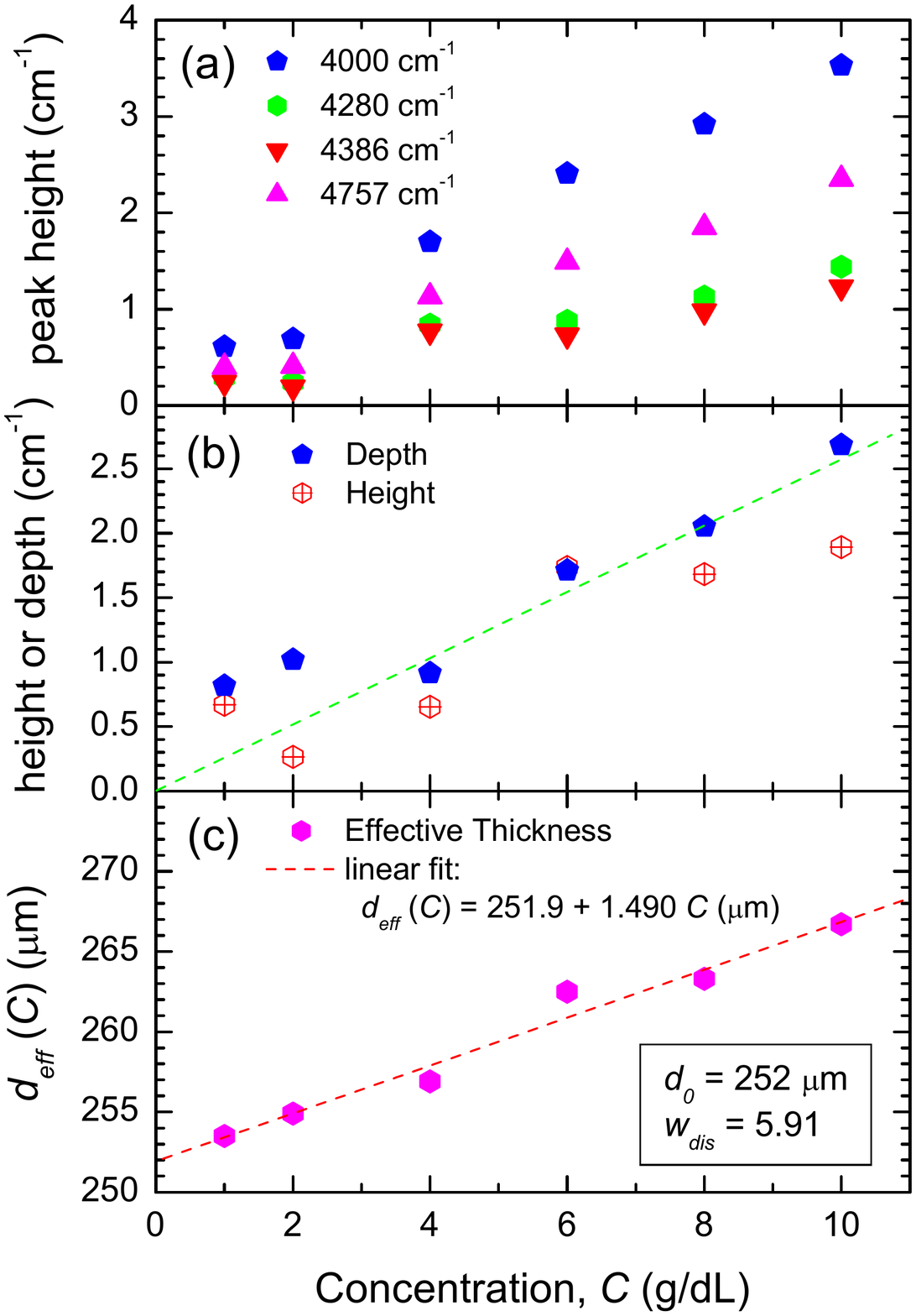}
  \caption{(Color online) (a) The concentration dependent peak height of four absorption modes of glucose in the combination region. (b) The concentration dependent height and depth of the new feature in figure \ref{fig1b} (b). (c) Concentration dependent effective thickness of the cell extracted from the water subtraction procedure (see in the text). We observe a strong linear relationship between the effective thickness of water and the glucose concentration.}
  \label{fig3}
\end{figure}

\begin{figure}[h]
  \centering\includegraphics[width=4.5 in]{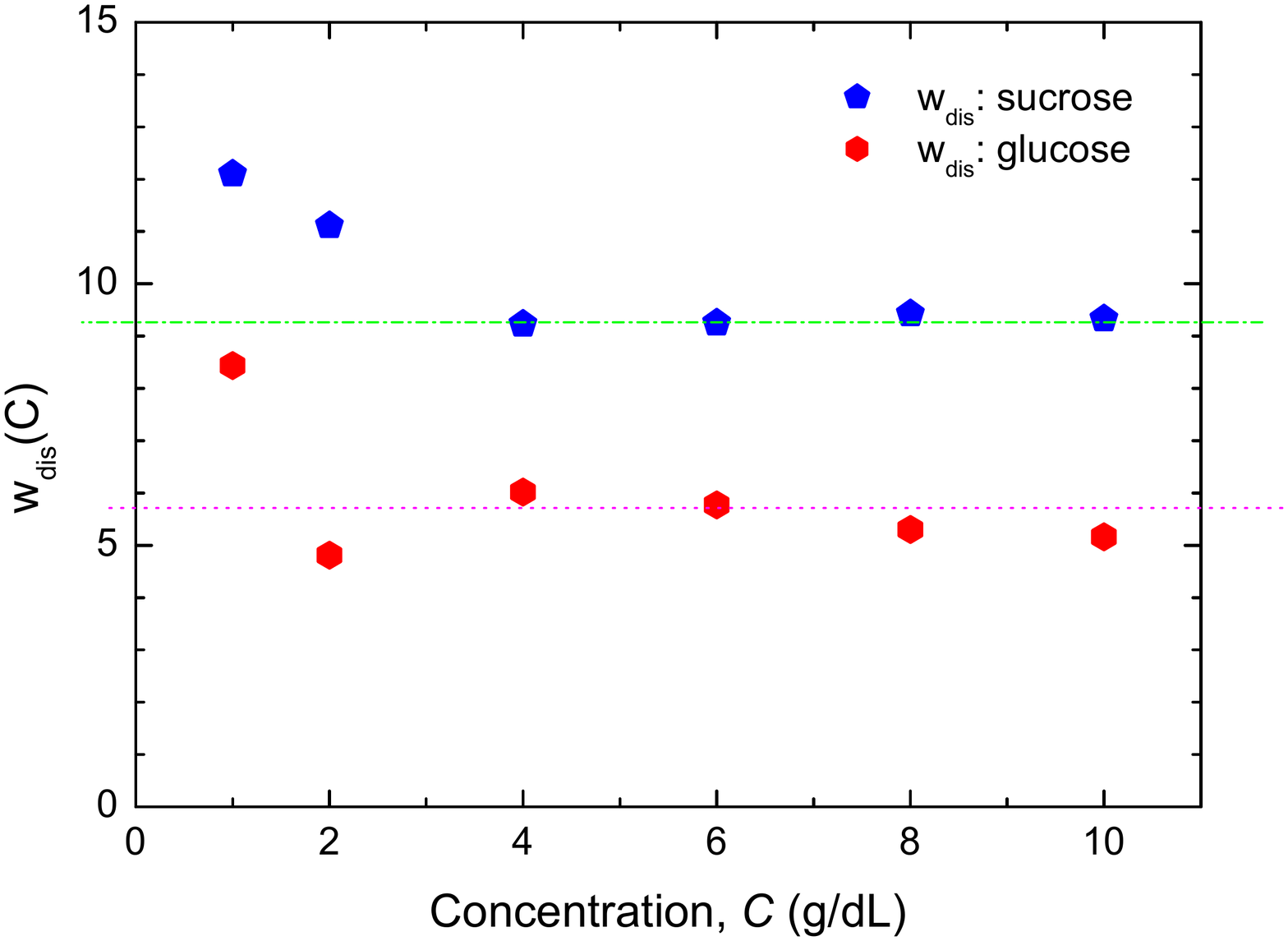}
  \caption{(Color online) We display the concentration dependent water displacement of glucose and sucrose obtained by using eq. \ref{eq9}.}
  \label{fig3b}
\end{figure}

Our approach for solving the downward peak problem in the difference spectra shown in figure \ref{fig1b} (a) is as follows. There is absence or very weak absorption of glucose around 5200 \cm, where water has a very strong absorption peak. It means that at that frequency absorption values of all solutions including pure water should be the same if we use a proper thickness of water for each solution. We performed the following procedure to remove an appropriate water absorption from the total absorption of each solution; by adjusting the thickness of pure water so that its absorption value at 5200 \cm\ is the same as the absorption value of each solution, we can then subtract a proper water spectrum from each solution spectrum to obtain pure glucose absorption due to glucose alone in each solution. We call the proper thickness of water for each solution the effective thickness of water. The procedure can be formulated as follows:
\begin{equation}\label{eq1b}
\alpha_{sol}(\omega, C)-\alpha_{w,d_{eff}}(\omega)=-\frac{\ln [T_{sol}(\omega, C)]}{d_0}-\Big{[}-\frac{\ln[ T_{w}(\omega)]}{d_{eff}(C)}\Big{]}
\end{equation}
where $d_{eff}(C)$ is the effective thickness of pure water for each concentration and $\alpha_{w,d_{eff}}(\omega)$ is the absorption coefficient of water calculated using the effective thickness $d_{eff}(C)$, which is dependent of the concentration. By performing the procedure, the concentration dependent negative peak, that was caused by subtracting too much water absorption, is removed although not completely. We will discuss this remaining downwards peak in the following paragraphs. This uncovers new absorption peaks due to glucose that agree with the pure glucose absorption shown in figure \ref{fig1a}, the dotted dark blue curve. The resulting absorption spectra due to glucose in six solutions are shown in figure \ref{fig1b} (b). In contrast to the single peak evidence in figure \ref{fig1b} (a) at 4700 \cm, three others are obvious in the combination region and two in the first overtone region. Also, the peak centered at 4700 \cm\ in figure \ref{fig1b} (a) has undergone a shift to the actual peak position of glucose at 4740 \cm. Glucose absorption bands in the first overtone region were also uncovered (see in the inset of figure \ref{fig1b} (b)), displaying absorption at 5700 and 6360 \cm. We note that the relative intensities of the glucose peaks are revealed as well. The concentration dependent peak heights of the four peaks in the combination region are displayed in figure \ref{fig3} (a). They show an almost a linear dependence on concentration.

In the inset of figure \ref{fig1b} (b) the peaks centered at 5700 and 6360 \cm\ have some visible discrepancies showing variation from the expected concentration dependence. In the first overtone region it seems that the 4 g/dL solution (green curve) has a larger absorption value than 6 g/dL (blue curve). However, looking at the 6 g/dL peak, it has a better-defined shape. Even though the 4 g/dL solution has a higher absorption value, the 6 g/dL solution has a more well-defined absorption compared to its average height in the first overtone region indicating a stronger real absorption. The same can be said for the 2 g/dL absorption (red curve) due to glucose, which appears to have a lower absorption than the 1 g/dL solution (black curve) in this region. Even though the area under the curves suggests that some weaker solutions have stronger absorption, the shape of the absorption peaks gives additional information about the concentration and a more accurate depiction of the concentration dependence in the first overtone region. These results can be attributed to the broader absorption peak of glucose in this region.\cite{amerov04} Due to a broader or worse defined peak, it is more difficult to detect proper concentration levels through aqueous media. A sharper or narrower peak provides a greater chance to see the concentration dependence of the absorption peak at that frequency because it is a better defined peak.

As we pointed out previously in figure \ref{fig1b} (b) we still have an extra feature near 5200 cm$^{-1}$, which has both a peak (on lower frequency side) and a dip (on higher frequency side). To understand this feature we simulate it with Lorentzian (reference) peaks. We found that there are three ways to produce such a feature. We show results of the three ways in figure \ref{fig2} (b), \ref{fig2} (c) and \ref{fig2} (d), respectively. In figure \ref{fig2} (a) we show the reference Lorentzian peak along with three peaks shifted  the horizontal axis by $\pm$ 5 \cm\ and one shifted peak by -100 \cm. The Lorentzian peak can be described as follows:
\begin{equation}\label{eq_Loren}
L_{peak}(\omega)=\frac{A}{\pi}\frac{\Gamma/2}{(\omega-\omega_c)^2+(\Gamma/2)^2}
\end{equation}
where $A$ is area under the peak, $\omega_c$ is the center frequency of the peak, and $\Gamma$ is the width of the peak, which is the full width at half maximum (FWHM). In figure \ref{fig2} (b) we show difference between each shifted peak with various shifting amounts ($\omega_{shift} =$ -100, -5, -4, -3, -2, -1 and +5 \cm) with a fixed width ($\Gamma =$ 400 \cm) and a fixed amplitude ($A =$ 200 cm$^{-2}$) and the reference peak; $L_{shift}(\omega)-L_{ref}(\omega)=(A/\pi)(\Gamma/2)
\{1/[(\omega-\omega_{shift}-\omega_c)^2+(\Gamma/2)^2]-1/[(\omega-\omega_c)^2+(\Gamma/2)^2]\}$. As we can see in the figure the more shifting produces the larger and better defined difference spectra. We note that the difference between the peak and dip positions do not change very much up to -100 \cm\ shift because of a large width 400\cm of the peak; 243 \cm\ for -100 case and 230 \cm\ for -5 case. But more shifting causes the larger interval between the peak and the dip. The frequency at the zero crossing is shifted by a half the frequency shift amount; the difference for -100 \cm\ case the zero crossing frequency is 1150 \cm. In figure \ref{fig2} (c) we show difference between each peak at 5195 \cm\ (i.e. $\omega_{shift}=$ 5 \cm) with various amplitudes ($A$ = 50, 100, 150 and 200 cm$^{-2}$) and the reference peak; $L_{amplitude}(\omega)-L_{ref}(\omega)=(A/\pi)(\Gamma/2)
\{1/[(\omega-\omega_{shift}-\omega_c)^2+(\Gamma/2)^2]-1/[(\omega-\omega_c)^2+(\Gamma/2)^2]\}$, where $\Gamma =$ 400 \cm. Here we also change the amplitude of the reference peak according to each peak amplitude. The results are shown in the figure; the more intense peaks give the larger differences. We note that the peak and dip positions are not at all dependent of the amplitude. In figure \ref{fig2} (d) we show the difference between each broadened peak at 5195 \cm\ (i.e. $\omega_{shift}=$ 5 \cm) with various widthes ($\Gamma^{\prime} =$ 400, 450, 500, 550, 600, and 650 \cm) and the reference peak; $L_{width}(\omega)-L_{ref}(\omega) = (A/\pi)\{(\Gamma^{\prime}/2)/[(\omega-\omega_{shift}-\omega_c)^2+(\Gamma^{\prime}/2)^2]
-(\Gamma/2)/[(\omega-\omega_c)^2+(\Gamma/2)^2]\}$, where $A =$ 200 cm$^{-2}$ and $\Gamma =$ 400 \cm. As we can see in the figure the sharpest peak gives the largest and most-defined difference. We also note that the peak (dip) position is red (blue) shifted as the width increases.

From observation of these three cases we can conclude that the extra feature in figure \ref{fig1b} (b) can be attributed to the second case i.e. amplitude changes with concentration. The intensity of the sharp absorption band edge due to the interacting water around 5200 \cm is getting larger as the amount of glucose increases, which is reasonable because more water molecules get involved in the interaction with the glucose molecules as the concentration increases. We do not expect a concentration dependent change in the frequency of the interacting water as long as we keep a relatively low glucose concentration in the solution.

\begin{figure}[h]
 \centering\includegraphics[width=4.5 in]{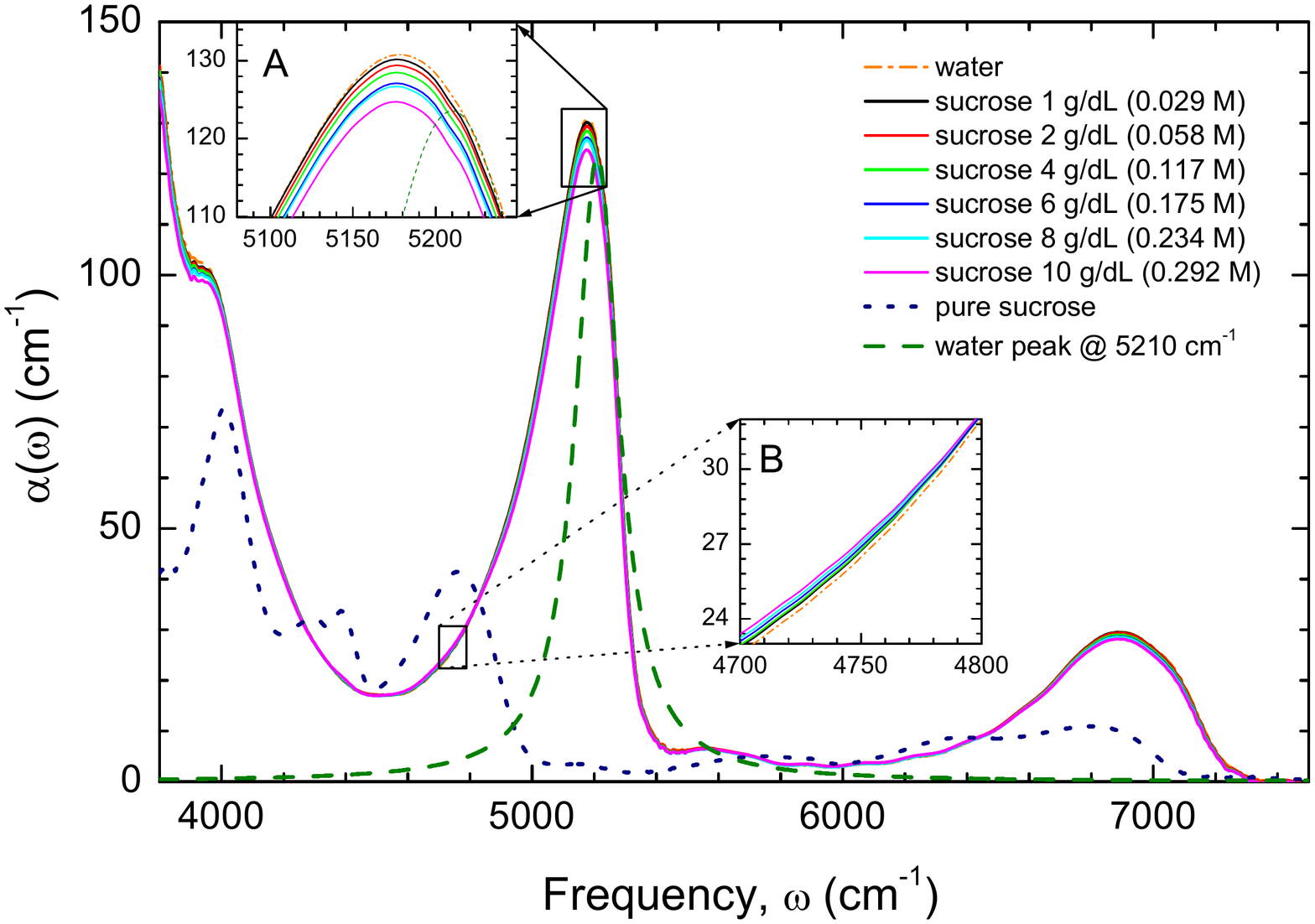}
  \caption{(Color online) Absorption coefficients of a pure amorphous sucrose (dotted dark blue) a pure water (dash-dotted orange), and six glucose aqueous solutions (from top to bottom; from low to high concentrations).  We also show a water peak at 5200 \cm. In the inset A and B we show expanded views near 5150 \cm\ and 4750 \cm\ respectively.}
  \label{fig4a}
  \end{figure}

  \begin{figure}[h]
\centering\includegraphics[width=3.5 in]{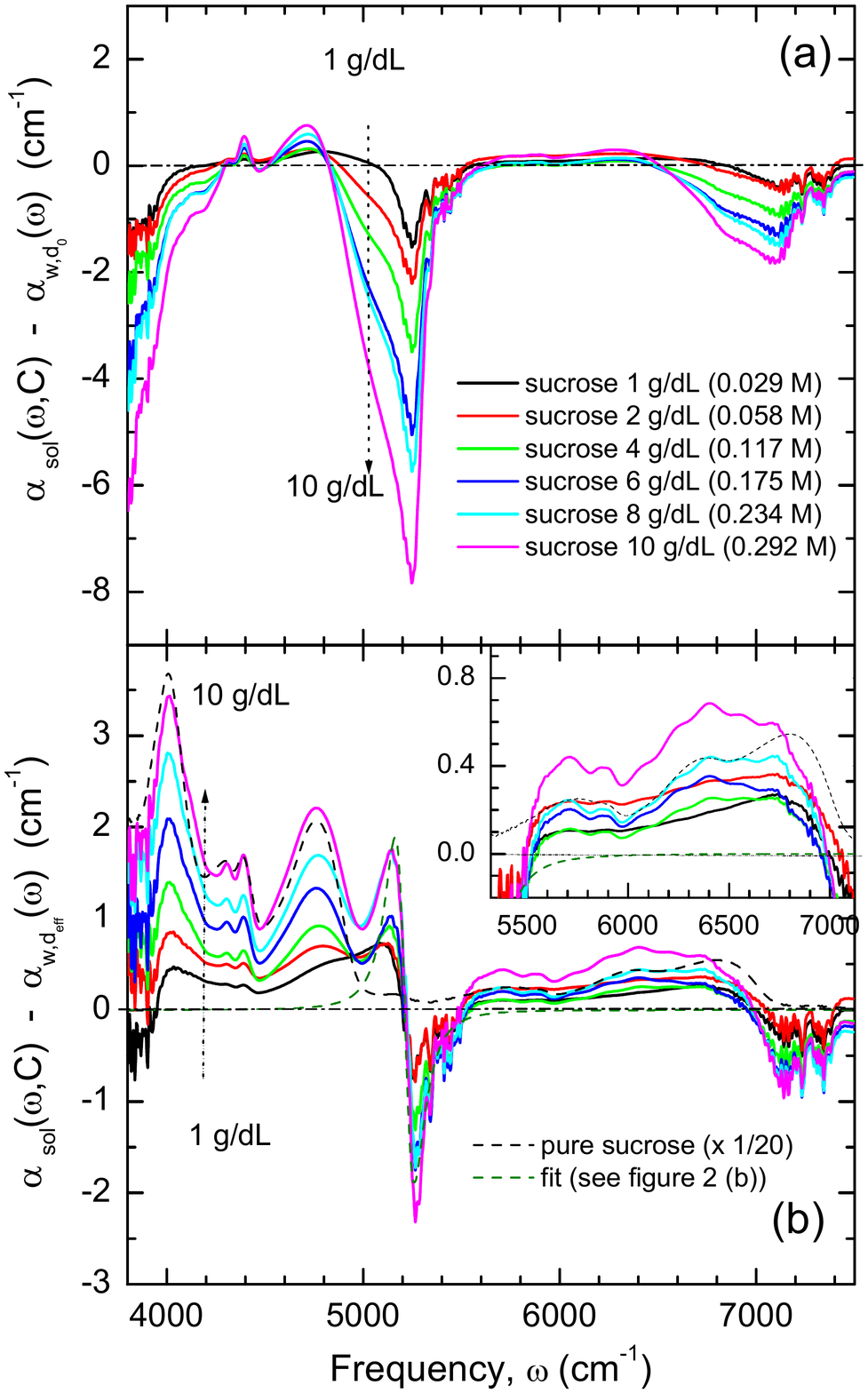}
  \caption{(Color online) (a) Sucrose absorption bands in solutions obtained at six different concentrations by subtracting water spectrum from those of the sucrose solutions. Water displacement coefficient of sucrose has not been considered on these spectra (see in the text). (b) Sucrose absorption coefficients in six different solutions. Water displacement effects have been accounted for the subtracting procedure (see in the text). The black dashed line in the lower frame is the pure glucose absorption spectra with its intensity reduced by a factor of 20. In the inset we display an expanded view to show spectral features better in the first overtone region.}
  \label{fig4b}
  \end{figure}

\begin{figure}[h]
  \centering\includegraphics[width=3.5 in]{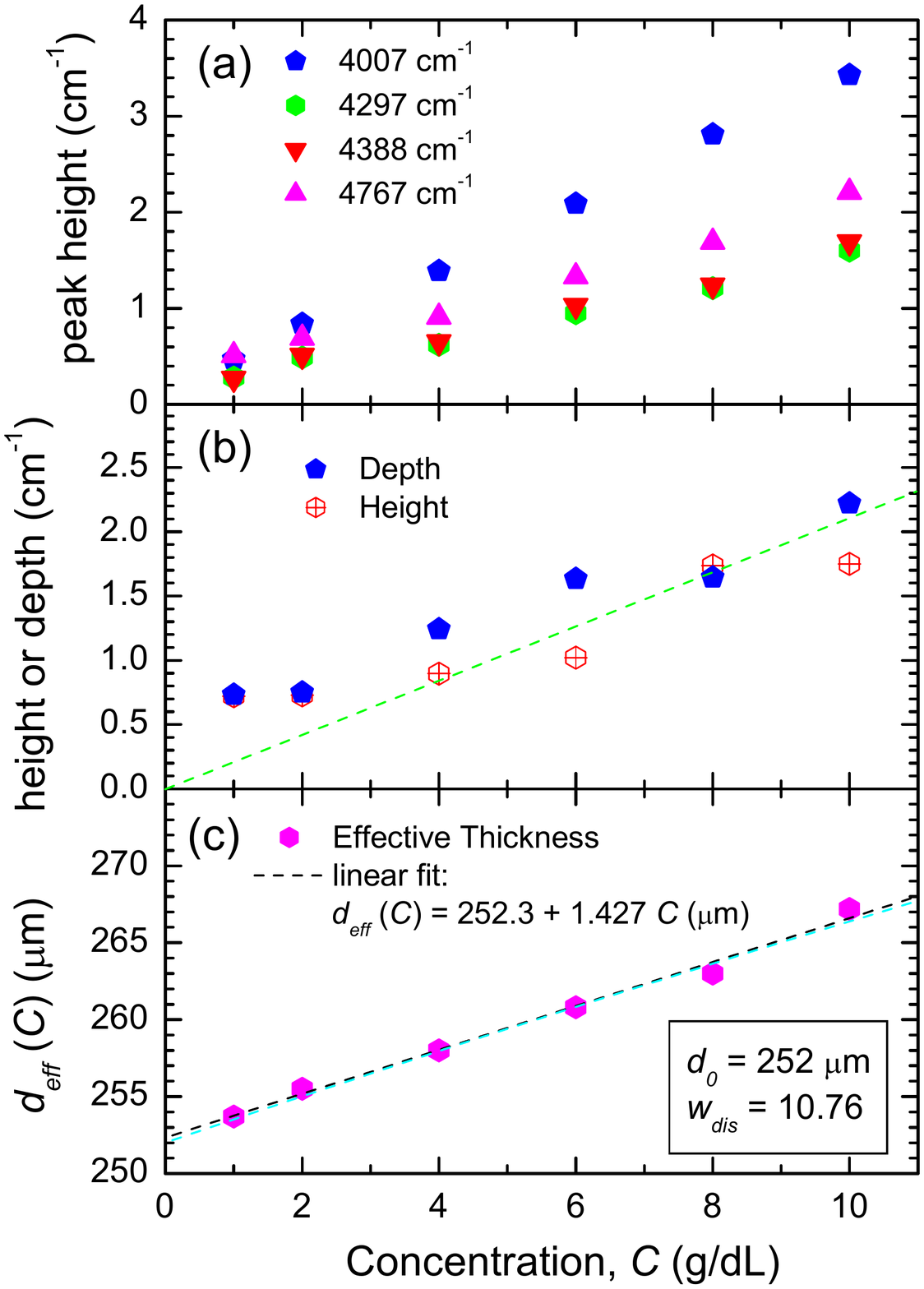}
  \caption{(Color online) (a) The concentration dependent peak height of four absorption modes of sucrose in the combination region. (b) The concentration dependent height and depth of the new feature in figure \ref{fig4b} (b). (c) Concentration dependent effective thickness of water. We observe a strong linear relationship between the effective thickness of water and the concentration.}
  \label{fig5}
\end{figure}

We fit the observed peak and dip feature near 5200 \cm\ in figure \ref{fig1b} (b) by using the model of our second case. In figure \ref{fig1a} we show the reference water peak at 5210 \cm\ (green dashed line), here we only considered the sharp absorption edge part of the water absorption band near 5200 \cm.\cite{nowak95} In figure \ref{fig1b} (b) we show an example fit (green dashed line) to the peak and dip feature near 5200cm\ in the 10 g/dL spectrum. We note that the amplitude (or area) of the reference peak is 3300 cm$^{-2}$ and its width is 170 \cm. From the fit we find that the amount of red shift is quit small, $\simeq$ 2 \cm. However, the resolution of the frequency shifting depends on the width of the reference peak considered. In our case the width (170 \cm) is quite large compared with the shift (2 \cm) so it is not very resolvable. Still what we can tell clearly is that the water absorption peak experiences a red shift. Here one may wonder that the instrumental resolution used to collect the spectra is 5 \cm\ whereas the observed shift in the water absorption band is around 2 \cm. However, we measure the resulting feature from the shift, which is much broader (about the width of the reference peak) than the instrumental resolution shown in the figure. We also expect to observe red shifts from other water peaks. To show this we display the absorption of pure glucose and the extracted glucose absorption spectrum, $\alpha_{sol}(\omega, C)-\alpha_{w,d_{eff}}(\omega)$ for $C=$ 10g/dL in a same panel as shown in figure \ref{fig3a}. In the figure we observe signatures of red shifts for other water peaks clearly. The signatures, which are strong dips, appear near sharp edges for water absorption namely, 3800 \cm, 5200 \cm, and 7000 \cm. So the red shifts seem to occur for all water absorption peaks. Roughly, we can tell that the height of the peak or depth of the dip of the feature can be a measure of the intensity of the interaction (see figure \ref{fig2} (c)). The resulting concentration dependent intensities of the red shift, height, and depth are displayed in figure \ref{fig3} (b). As we expected, the extracted intensity is roughly proportional to the glucose concentration. The deviation from the linearity may come from the uncertainty in the water subtraction procedure.

From the appropriate water subtraction procedure, which we described previously, we can obtain the effective thickness of water for each glucose solution. Figure \ref{fig3} (c) displays the extracted effective thickness of water as a function of the glucose concentration. It shows a strong linear relationship between the effective thickness and the concentration from 1 g/dL through 10 g/dL. The water displacement coefficient is defined by the molar concentration change of water caused by the dissolution of a unit molar concentration of the solute. The molar concentration is defined by the number of moles per a liter of solvent (in our case, water). More practically, the water displacement coefficient is the number of water molecules which are replaced by a solute molecule in the solution. By using these definitions we can write down the effective thickness of water in the cell as a function of glucose concentration.
\begin{eqnarray}\label{eq2}
d_{eff}(C) &=&  d_{0}\Big{[} 1+ \frac{C^{\prime}\cdot w_{dis}}{1+C^{\prime} \cdot w_{dis}}\Big{]} \:\:\: \mbox{and} \nonumber \\
C^{\prime} &\equiv& \frac{ M_{water}}{100 \cdot M_{solute}}C
\end{eqnarray}
where $C$ is the concentration in g/dL, $d_{eff}(C)$ is the concentration dependent effective thickness of water, $M_{water}$ is the molecular weight of water,  $M_{solute}$ is the molecular weight of solute, $d_{0}$ is the real thickness of the cell (in our case, 252 $\mu$m) and $w_{dis}$ is the water displacement coefficient. We see that $C^{\prime}$ is a small quantity; 0.01 for 10 g/dL glucose solution and 0.0053 for 10 g/dL sucrose solution, these are the maximum values for glucose and sucrose solutions. The water displacement coefficient is roughly a single digit value. So $C^{\prime}\cdot w_{dis}$ is small and we can rewrite the equation (\ref{eq2}) approximately as follows:
\begin{eqnarray}\label{eq2a}
d_{eff}(C)&\cong&d_0[1+ C^{\prime}\cdot w_{dis}\cdot (1-C^{\prime}\cdot w_{dis})]\:\:\: \mbox{or} \nonumber \\
d_{eff}(C)&\simeq&d_0[1+C^{\prime}\cdot w_{dis}]  \nonumber \\
&=&d_0+\frac{d_0\cdot M_{water} \cdot w_{dis}}{100 \cdot M_{solute}}\:C
\end{eqnarray}
In the lower equation we made a further approximation assuming $C^{\prime}\: w_{dis} \ll 1$. This last equation shows that $d_{eff}(C)$ is linear in the concentration $C$, which is what we obtain from our analysis and the results shown in figure \ref{fig3} (c). We fit the data points in figure \ref{fig3} (c) to a straight line; $d_{eff}(C)=251.9 \:(\pm \:0.8) + 1.490 \:(\pm\: 0.128) \:C \:\: (\mbox{in $\mu$m})$. From the linear fitting we obtain the real thickness of the cell and the slope of the straight line. By using the fitting parameters we are able to estimate the water displacement coefficient of glucose.
\begin{equation}\label{eq3}
w_{dis}=slope\:\frac{M_{solute}\cdot 100}{M_{water}\cdot d_0}
\end{equation}
where $slope$ is the slope of the straight line. The water displacement coefficient of glucose obtained is $ 5.91 \pm 0.51$ at 23 $^\circ$C. This means that one glucose molecule can take up the space of 5.91 water molecules in the solution. This value seems to be rather large compared to a value of 5.051 at 21 $^\circ$C reported in Ref [23]. We also note that the relative standard deviation for the proposed method is 8.6 \% (0.51/5.051) compared to a value of 0.056 \% (0.0035/6.245) from the direct density method reported in Ref [14]. The precision of this method for obtaining the water displacement coefficient is not as good as that of the density method. A new method introduced in the following paragraph allows us to obtain a concentration dependent water displacement coefficient. The concentration dependent water displacement coefficient of glucose (see lower figure \ref{fig3b}) shows a better value for the coefficient at high concentrations; for 10 g/dL sample, the water displacement coefficient is 5.10, which seems to be a more reliable value for the coefficient at 23 $^\circ$C.

We also can obtain the water displacement coefficient from an another simpler method as follows. By comparing intensities of an absorption peak of pure glucose (for example, at 4000 cm$^{-1}$) in figure \ref{fig1a} and the corresponding peak of glucose alone in solution in figure \ref{fig1b} (b) we can estimate an effective thickness of glucose alone in solution compared to the total thickness of the solution. If the total volume of the solution consists of glucose alone the absorption coefficient of the solution sample would be identical to that of pure glucose. Since the absorption intensity of glucose is proportional to the effective thickness (or amount) of glucose alone in solution the intensity ratio is the same as the effective thickness ratio as the following equation,
\begin{equation}\label{eq8}
\frac{A_{solute.sol}}{A_{solute.pure}}=\frac{C/M_{solute}\cdot w_{dis} \cdot N_{A}}{[100/M_{water}+C/M_{solute}\cdot w_{dis}] \cdot N_{A}}
\end{equation}
where $A_{solute.pure}$ and $A_{solute.sol}$ are the peak heights of pure glucose and glucose in solution, respectively, $C$ is the concentration in g/dL and $N_{A}$ is the Avogadro's number. When we solve for the water displacement coefficient, $w_{dis}$, we get the following equation.
\begin{equation}\label{eq9}
w_{dis}=\frac{A_{solute.sol}\cdot 100 \cdot M_{solute}}{(A_{solute.pure}-A_{solute.sol})\cdot C\cdot M_{water}}
\end{equation}
This equation means that for a given concentration, if we know the water displacement of solute and its absolute absorption coefficient we can easily estimate the absorption coefficient of solute alone in solution. In other word if we can measure the absorption coefficient of solute alone in solution for a given concentration we are able to obtain the water displacement coefficient of the solute. For example, we consider 10 g/dL glucose solution and the absorption peak at 4000 cm$^{-1}$. Then by using eq. (\ref{eq9}) $w_{dis}=(3.52\times 100 \times 180)/[(72.5-3.52)\times 10\times 18] \cong$ 5.10. Even though the value is slightly smaller than the previously extracted value (5.91) it is consistent with the previous one. The concentration dependent water displacement of glucose is shown in Fig. \ref{fig3b} along with the concentration dependent water coefficient of sucrose. It seems to be independent of concentration even though there are some noisy data points in the low concentration region due to the uncertainty in the water subtraction procedure.

The water displacement coefficient of glucose shows a strong temperature dependence; 5.051 at 21 $^\circ$C~\cite{kohl95} and 6.245 at 37 $^\circ$C~\cite{amerov04} and our extracted water displacement coefficients of glucose seem to be consistent with other studies; at least our value is in between those two values obtained at two lower and higher temperatures. The temperature dependence in the water displacement coefficient probably comes mostly from thermally induced morphology change of glucose molecules in the water. More systematic studies should be performed on temperature dependent water displacement of glucose. We note that, at very high concentrations, the linear trend can not hold because interaction between nearest glucose molecules which is an indirect repulsive interaction in water will become stronger as the concentration increases.

We performed the same experiment and data analysis with a different solute, sucrose (C$_{12}$H$_{22}$O$_{11}$), which has a larger molecular weight. The measured data and analysis results are displayed in figure \ref{fig4a}, \ref{fig4b} and figure \ref{fig5}. Although there are some detailed qualitative differences, the overall qualitative concentration dependent trends are very similar to those of glucose. There are four sucrose absorption peaks in the combination region and two peaks in the first overtone region as for glucose. One thing which we would like to note is that the relative intensities between the peaks are different. As we can see in the pure sucrose absorption spectrum, the peak at 4390 \cm\ is relatively large. The recovered concentration dependent peak height of sucrose absorption modes in the combination region are displayed in figure \ref{fig5} (a). They show a linear dependence on the concentration as we expected. We observe an additional feature other than sucrose absorption peaks in figure \ref{fig4b} (b) as those in figure \ref{fig1b} (b). We show an example fit to the feature for 10 g/dL only for the absorption edge region. From this fitting we get that the red shift is very small around 2 \cm. The concentration dependent height and depth of the feature are shown in figure \ref{fig5} (b). This indicates that water molecules around a sucrose molecule is not free; the absorption peaks due to the water will be shifted to lower frequencies, i.e. red shifts, which is due to water structure enhancement within the hydration shells around solute molecules.\cite{hvidt83,raschke05}

We also display the concentration dependent effective thickness, $d_{eff}(C)$, of water in the liquid cell in figure \ref{fig5} (c). We fit the data to a straight line; $d_{eff}(C)=252.3\: (\pm\:0.4) + 1.427\: (\pm\: 0.065) \:C \:\: (\mbox{in $\mu$m})$. By using eq. (\ref{eq3}), and the extracted slope of the line and $d_0$ from the fit we can get the water displacement coefficient of sucrose, $ 10.76 \pm 0.46$ at 23 $^{\circ}$C. The water displacement coefficient tells us that a sucrose molecule can take up space equivalent to 10.76 water molecules in the solution. We also used the other method (see eq. (\ref{eq9})) to obtain the water displacement coefficient of sucrose. For example, we consider the 10g/dL sucrose solution and the absorption peak at 4000 cm$^{-1}$. Then by using eq. (\ref{eq9}), $w_{dis}=(3.57\times 100 \times 342)/[(72.0-3.57)\times 10\times 18] \cong$ 9.35. Even though the value is slightly smaller than the previously extracted value (10.76) it is quite consistent with the previous one. The concentration dependent water displacement of sucrose is shown in Fig. \ref{fig3b}. In low concentration region the data show more noise and seem to deviate from the concentration independent trend at high concentration due to the uncertainty in the water subtraction procedure in the low concentration region. The molecular weight of sucrose ($M_{sucrose}$ = 342) is almost twice of glucose ($M_{glucose}$ = 180) i.e. in solution a sucrose molecule also takes up almost twice the space of a glucose molecule. This may indicate that sucrose molecules in solution have elongated shapes instead of global ones.\\

{\noindent}\textbf{CONCLUSIONS}

It is clear from our work as well as previous studies that water displacement is an important quantity that must be considered for a proper study of glucose concentration in aqueous media and blood. Our new method is fundamentally different from the study of Amerov, Chen, and Arnold, who realized the problem introduced by water displacement and compensated for it by using an independently measured water displacement coefficient based on density measurements.\cite{amerov04} We introduce a new spectroscopic method which is simple where we remove water absorption bands accurately from measured glucose aqueous solutions without independent measurement of water displacement by the glucose. Using the linear relationship on concentration found for the effective thickness of water in the liquid cell we can properly manipulate the spectra to remove water absorption and obtain reliable concentration dependent glucose absorption bands. In the spectra obtained from the subtraction procedure we observed signatures of interaction between water and solute molecules in the solution. The red shift of the water absorption near 5200 \cm\ is around 2 \cm, which is very small but clearly observable in the spectra. As we mentioned previously, the amount of shifting is not easily resolvable since the width of water absorption is quite broad compared with the red shift. Additionally, we were able to extract the water displacement coefficient of glucose, which is consistent with values reported in literatures.\cite{amerov04,kohl95} These results may help to monitor non-invasively the glucose level in human body.

This method has several advantages: first of all, by removing the need to measure the water displacement coefficient independently, we can estimate a reliable water displacement coefficient from our concentration dependent spectrum, making it a self-consistent method which we can apply to other solutions. Actually, we also applied the same method to sucrose aqueous solution and got a reasonable water displacement coefficient of sucrose. The method of water displacement extraction in this paper can provide as a new general method using optical spectroscopy technique for other biological or organic materials. This method also allows us to observe red shifts in the difference spectra, which is not easy to detect with different experimental techniques. Similar red shifts were observed by other groups and other experimental techniques in different aqueous solutions, or water-oil and water-air interfaces. \cite{moore06,perera09} Our approach provides another useful tool to study water molecules in the hydration cell around a solute molecule in aqueous solutions.\\

{\bf Acknowledgement} We thank T. Timusk and R. Peters for useful discussions. This work has been supported by the special fund of Department of Physics at Pusan National University, Busan, Republic of Korea. This work also was supported by the National Research Foundation of Korea Grant funded by the Korean Government (NRF-2010-371-B00008).

\newpage
{\noindent}\textbf{List of Figure Captions}\\

{\noindent}Fig. 1. Absorption coefficients of a pure amorphous glucose (dotted dark blue line) a pure water (dash-dotted orange line), and six glucose aqueous solutions (from top to bottom; from low to high concentrations). We also show a water peak at 5200 \cm. In the inset A and B we show expanded views near 5150 \cm\ and 4750 \cm\ respectively.\\

{\noindent}Fig. 2. (a) Glucose absorption bands in solutions obtained at six different concentrations by subtracting water spectrum from those of the glucose solutions. Water displacement coefficient of glucose has not been considered on these spectra (see in the text). (b) Glucose absorption coefficients in six different solutions. Water displacement effects have been taken into account for the subtracting water absorption procedure (see in the text). The green dotted line is a fitted line to the feature near 5200 \cm with both peak and dip from a model calculation (see figure \ref{fig2} (b) and corresponding text). The black dashed line in the lower frame is the pure glucose absorption spectra with its intensity reduced by a factor of 22. In the inset we display an expanded view to show spectral features better in the first overtone region.\\

{\noindent}Fig. 3. (a) We show a reference peak which has its center at 5200 \cm\ and width 400 \cm. We also show two shifted peaks in horizontal direction by negative 5 \cm\ (dashed dotted red curve) and positive 5 \cm\ (dashed blue curve), respectively. In the inset we expand the graphs near the peak region to show the shifts more clearly. (b) We also show resulting differences subtracted the reference peak at 5200 \cm\ from the shifted peaks by seven different amounts  (see in the text). (c) We show resulting differences subtracted the reference peak at 5200 \cm\ from the peaks at 5205 \cm\ with four different amplitudes (see in the text). (d) (c) We show resulting differences subtracted the reference peak at 5200 \cm\ from the peaks at 5205 \cm\ with six different widths (see in the text).\\

{\noindent}Fig. 4. Comparison the pure glucose absorption and the extracted glucose absorption ($\alpha_{sol}(\omega, C)-\alpha_{w,d_{eff}}(\omega)$) from 10g/dL solution. We note that we reduce the intensity of the pure glucose absorption by a factor of 22. We also show the water spectrum as well for comparison purpose and reduce its intensity by a factor of 40.\\

{\noindent}Fig. 5. (a) The concentration dependent peak height of four absorption modes of glucose in the combination region. (b) The concentration dependent height and depth of the new feature in figure \ref{fig1b} (b). (c) Concentration dependent effective thickness of the cell extracted from the water subtraction procedure (see in the text). We observe a strong linear relationship between the effective thickness of water and the glucose concentration.\\

{\noindent}Fig. 6. We display the concentration dependent water displacement of glucose and sucrose obtained by using eq. (\ref{eq9}).\\

{\noindent}Fig. 7. Absorption coefficients of a pure amorphous sucrose (dotted dark blue) a pure water (dash-dotted orange), and six glucose aqueous solutions (from top to bottom; from low to high concentrations).  We also show a water peak at 5200 \cm. In the inset A and B we show expanded views near 5150 \cm\ and 4750 \cm\ respectively.\\

{\noindent}Fig. 8. (a) Sucrose absorption bands in solutions obtained at six different concentrations by subtracting water spectrum from those of the sucrose solutions. Water displacement coefficient of sucrose has not been considered on these spectra (see in the text). (b) Sucrose absorption coefficients in six different solutions. Water displacement effects have been accounted for the subtracting procedure (see in the text). The black dashed line in the lower frame is the pure glucose absorption spectra with its intensity reduced by a factor of 20. In the inset we display an expanded view to show spectral features better in the first overtone region.\\

{\noindent}Fig. 9. (a) The concentration dependent peak height of four absorption modes of sucrose in the combination region. (b) The concentration dependent height and depth of the new feature in figure \ref{fig4b} (b). (c) Concentration dependent effective thickness of water. We observe a strong linear relationship between the effective thickness of water and the concentration.

\end{document}